\documentclass[12pt]{article}
\usepackage{graphicx}

\newcommand{\h}[1]{\mathop{\lambda}\limits_{#1}\ \!\!\!}

\newcommand{\edf}{\ {\mathop{=}\limits^{\rm def}}\ }

\newcommand{\al}{\alpha}

\begin{document}
 \begin{center}
\bf { A PURE GEOMETRIC APPROACH TO STELLAR STRUCTURE}
\end{center}
\begin{center}
{\bf M. I. Wanas} \footnote{Astronomy Department, Faculty of Science,
Cairo University, Giza, Egypt.

E-mail:mwanas@cu.edu.eg; mamdouh.wanas@bue.edu.eg}$^{,~3,~4}$ $~\& ~$ {\bf Samah A. Ammar}~\footnote{Mathematics Department,
Faculty of Girls, Ain Shams University, Cairo, Egypt.

E-mail:samahammar@eun.eg}$^{,}$ \footnote{ Egyptian Relativity Group (ERG),
URL:www.erg.eg.net.\\
$~~~~~^4$Center for Theoretical Physics (CTP) at the British University in Egypt (BUE), Egypt.}

\end{center}
\begin{abstract}
The present work represents a step to deal with stellar structure
using a pure geometric approach. A geometric field theory is used
to construct a model for a
spherically symmetric configuration. The model obtained can be
considered as a pure geometric one in the sense that the tensor
describing the material distributions is not a phenomenological
object, but a part of the geometric structure used. A general
equation of state is obtained from, and not imposed on, the model.
The solution obtained shows that there are different zones
characterizing the configuration: a central radiation dominant zone,
a probable convection zone as a physical interpretation of the singularity
of the model and a corona like zone. The model may represent a type of  main
sequence stars. The present work shows that Einstein's geometerization
scheme can be extended to gain more physical information within
material distribution, with some advantages.

\end{abstract}
\newpage

\section{Introduction}

Gravity plays an important role in the structure and evolution of
macroscopic objects. It affects many physical parameters of such
objects, e.g. density, pressure and temperature. In relativistic
theories of gravity, especially general relativity (GR), the
relation between gravity and such parameters are given by the field
equations of the theory. In the case of GR, the field equations
within the material distribution are given by Einstein (1955)
$$R_{\mu\nu}-\frac{1}{2} g_{\mu\nu}R= -\kappa
T_{\mu\nu}.\eqno{(1.1)}$$ where $T_{\mu\nu}$ is a tensor giving the
physical properties of the material distribution, while the L.H.S.
of this equation gives the geometric description of the
gravitational field. In orthodox GR, written in the context of
Riemannian geometry, the L.H.S. of (1.1) is a pure geometric object
while the R.H.S. is a phenomenological one.

It is well known that when using (1.1) to study problems concerning
structure and evolution of large scale structures, one needs to impose
an extra condition ({\it an equation of state}) in order to solve
the problem. This is done because the field equations (1.1) are, in
general, not sufficient to get the unknown functions of the problem.
This scheme, of imposing an equation of state, leads some times to
unsatisfactory results (e.g. Schwarzschild interior solution  (cf. Adler, Bazin \& Schiffer 1975).
The solution, in this case, can be considered as a mathematical one,
without any physical applications. Reasons of this problem may be
implied by the R.H.S. of (1.1), which is not a part of the geometric
structure used, the Riemannian geometry. It is preferable, if we
use a theory in which the material-energy tensor is a part of the
geometric structure used. Since Riemannian structure is just
sufficient to describe gravitational fields, one needs a more wider
geometric structure, than the Riemannian one, in order to describe the material distribution.
Among the advantages of using a pure geometric model are:\\
(i) The equation of state, characterizing the material distribution within the model, is predicted by the model and not imposed from outside. In this case one don't need an extra condition (arbitrary) to solve the field equations.
(ii) The scheme will show that matter can be induced by geometry. This would be similar to the case of Schwarzschild exterior solution, in which the source of the field (gravitational mass) is induced by geometry.

The aim of the present work is to use a pure geometric field theory,
i.e. a theory in which the tensor describing the material-energy
distribution is a part of the geometric structure, in applications.
For this reason we give, in section 2, a brief review of a geometric
structure, with simultaneously non-vanishing curvature and torsion.
In section 3 we review briefly the field equations of a pure
geometric theory written in the context of the structure given in
section 2. In section 4 we solve these field equations in the case
of spherical symmetry, restricting it to a special case of a perfect
fluid. In section 5, we discuss the model obtained. Some concluding
remarks are given in section 6.
\section{ The Underlying Geometry}

In the present section we are going to review, briefly, a geometric
structure, more wider than the Riemannian one. This structure is a
version of "Absolute Parallelism" (AP)-geometry ( cf. Mikhail 1962), in which
both curvature and torsion are simultaneously non-vanishing objects
(Wanas 2001). The structure of the conventional AP-geometry is defined
completely (in 4-dimensions) by a tetrad vector field $\h{i}_\mu $
such that, its determinant $\lambda^{*} = \| \h{i}_{\mu} \|$ is non-
vanishing and,
$$ \h{i}^\mu \h{i}_\nu = \delta^{\mu}_{\nu}.
\eqno{(2.1)}
 $$ where $\h{i}^{\mu}$ are the contravariant components of
 $\h{i}_{\mu}$. Einstein's summation convention is carried over
 repeated indices wherever they exist.
 Using these vectors, one can define the second order symmetric
tensor,
 $$g_{\mu\nu}\edf \h{i}_{\mu} \h{i}_{\nu},\eqno{(2.2)}$$
 which is clearly non-degenerate. This symmetric tensor can be used to play the role of the metric of Riemannian
space, associated with the AP-space, when needed. \\

\underline{\bf Connections, Curvatures and Torsion:}

The AP-space admits a non-symmetric affine connection
~~$\Gamma^{\al}_{~~\mu\nu}$, which arises as a consequence of the
AP-condition (cf. Mikhail 1962), i.e. $$\h{i}_{\stackrel{\mu}{+}|~
\nu}\edf\h{i}_{\mu,\nu}-\Gamma^{\al}_{.~\mu\nu}\h{i}_{\al}=0,
\eqno{(2.3)} $$ where the stroke and the (+) sign denote tensor
differentiation as defined \footnote { we are going to use (,) for
ordinary partial differentiation and (;) to represent covariant
differentiation using Christoffel symbol.}. Equation (2.3) can be
solved to give,
$$\Gamma^{\al}_{.~\mu\nu}=\h{i}^{\al}\h{i}_{\mu,\nu}.\eqno{(2.4)}$$
Since $\Gamma^{\al}_{.~\mu\nu}$ is non-symmetric, as clear from
(2.4), then one can define the dual connection (cf. Mikhail 1962) as,
$$\tilde{\Gamma}^{\al}_{.~\mu\nu}\edf\Gamma^{\al}_{.~\nu\mu}.\eqno{(2.5)}$$
Also, Christoffel symbol $\{^{\al}_{\mu\nu}\}$ could be defined
using the metric tensor (2.2). So, in the AP-geometry one could
define four linear connections, at least: The non-symmetric
connection (2.4), its dual (2.5), Christoffel symbol defined using
(2.2), and the symmetric part of (2.4), $\Gamma^{\al}_{.(\mu\nu)}$.

The curvature tensors, corresponding to the above
mentioned connections, respectively, are (Wanas 2001)
$$M^{\al}_{.~\mu\nu\sigma}\edf\Gamma^{\al}_{.~\mu\sigma,\nu}-\Gamma^{\al}_{.~\mu\nu,\sigma}
+\Gamma^{\epsilon}_{.~\mu\sigma}\Gamma^{\al}_{.~\epsilon\nu}-\Gamma^{\epsilon}_{.~\mu\nu}
\Gamma^{\al}_{.~\epsilon\sigma},\eqno{(2.6)}$$ $$
\tilde{M}^{\al}_{.~\mu\nu\sigma}\edf
\tilde{\Gamma}^{\al}_{.~\mu\sigma,\nu}-
\tilde{\Gamma}^{\al}_{.~\mu\nu,\sigma}+
\tilde{\Gamma}^{\epsilon}_{.~\mu\sigma}\tilde{\Gamma}^{\al}_{.~\epsilon\nu}-\tilde{\Gamma}^{\epsilon}_{.~\mu\nu}
\tilde{\Gamma}^{\al}_{.~\epsilon\sigma},\eqno{(2.7)}$$
$$R^{\al}_{.~\mu\nu\sigma}\edf\{^{\al}_{\mu\sigma}\}_{,\nu}-\{^{\al}_{\mu\nu}\}_{,\sigma}
+\{^{\epsilon}_{\mu\sigma}\}\{^{\al}_{\epsilon\nu}\}-\{^{\epsilon}_{\mu\nu}\}
 \{^{\al}_{\epsilon\sigma}\},\eqno{(2.8)}$$ $$ \bar{M}^{\al}_{.~\mu\nu\sigma}\edf \Gamma^{\al}_{.~(\mu\sigma),\nu}
 -\Gamma^{\al}_{.~(\mu\nu),\sigma}+\Gamma^{\epsilon}_{.~(\mu\sigma)}\Gamma^{\al}_{.~(\epsilon\nu)}-\Gamma^{\epsilon
 }_{.~(\mu\nu)}\Gamma^{\al}_{.~(\epsilon\sigma)}.\eqno{(2.9)}$$ Due to the AP-condition
(2.3) the curvature tensor given by (2.6) vanishes identically,
while those given by (2.7), (2.8) and (2.9) do not vanish.

Using the non-symmetric connection one can define a third order
skew-tensor, $$\Lambda^{\alpha}_{.~\mu \nu} \edf
\Gamma^{\alpha}_{.~\mu \nu} - \Gamma^{\alpha}_{.~\nu \mu}=
\tilde{\Gamma}^{\al}_{.~\nu\mu}-\tilde{\Gamma}^{\al}_{.~\mu\nu}
=-\Lambda^{\al}_{.~\nu\mu}.\eqno{(2.10)}$$ This tensor is the {\it
torsion} tensor of AP-space. One can define another third order
tensor viz,
$$\gamma^{\al}_{.~\mu\nu}\edf\h{i}^{\al}\h{i}_{\mu;\nu}.\eqno{(2.11)}$$This
tensor is called the {\it contortion} of the space, using which, it
can be shown that
$$\Gamma^{\alpha}_{.~\mu\nu}=\gamma^{\al}_{.~\mu\nu}+\{^{\alpha}_{\mu\nu}\},
\eqno{(2.12)}$$ where $\gamma_{\mu\nu\al}$ is skew-symmetric in its
first two indices. A {\it basic vector} could be obtained by
contraction, using any one of the above third order tensors, $$
C_{\mu} \edf \Lambda^{\alpha}_{.~\mu \alpha }= \gamma^{\alpha}_{.~
\mu \alpha}. \eqno{(2.13)}$$ Using the contortion, we can define the
following symmetric third order tensor,
$$\Delta^{\al}_{.~\mu\nu}\edf\gamma^{\al}_{.~\mu\nu}+\gamma^{\al}_{.~\nu\mu}.
\eqno{(2.14)}$$ The skew-symmetric and symmetric parts of the tensor
$\gamma^{\al}_{.~\mu\nu}$ are, respectively
$$\gamma^{\al}_{.~[\mu\nu]}=\Gamma^{\al}_{.~[\mu\nu]}=\frac{1}{2}\Lambda^{\al}_{.~\mu\nu},\eqno{(2.15)}$$
$$\gamma^{\al}_{.~(\mu\nu)}=\frac{1}{2}\Delta^{\al}_{.~\mu\nu},\eqno{(2.16)}$$
where the brackets $[~]$ and the parenthesis $(~)$ are used for
anti-symmetrization and symmetrization, of tensors, with respect
to the enclosed indices, respectively.

\underline{\bf Tensor Derivatives}

Using the connections mentioned above, one can define the following
derivatives (Wanas 2001),
$$A^\mu_{+|~ \nu}\edf
A^{\mu}_{~,\nu}+\Gamma^{\mu}_{~\al\nu}A^{\al},\eqno{(2.17)}$$
$$A^\mu_{-|~ \nu}\edf
A^{\mu}_{~,\nu}+\tilde{\Gamma}^{\mu}_{~\al\nu}A^{\al},\eqno{(2.18)}$$
$$A^{\mu}_{~;\nu}\edf
A^{\mu}_{~,\nu}+\{^{\mu}_{\al\nu}\}A^{\al},\eqno{(2.19)}$$ $$
A^{\mu}_{~|~\nu}\edf A^{\mu}_{~,\nu}+ \Gamma^{\mu}_{.~(\al\nu)}
A^{\al}.\eqno{(2.20)}$$

The following table is extracted from Mikhail (1962) and contains second order
tensors that are used in most applications.
\newpage
\begin{center}
 Table 1: Second Order World Tensors (Mikhail 1962)     \\
\vspace{0.5cm}
\begin{tabular}{|c|c|} \hline
 & \\
Skew-Symmetric Tensors                &  Symmetric Tensors   \\
 & \\ \hline
 & \\
${\xi}_{\mu \nu} \edf \gamma^{~ ~ \alpha}_{\mu \nu .
|{\stackrel{\alpha}{+}}} $ &
\\

${\zeta}_{\mu\nu} \edf C_{\alpha}~{\gamma^{~~ \alpha}_{\mu \nu .} }
$ &
\\
 & \\ \hline
 & \\
${\eta}_{\mu \nu} \edf C_{\alpha}~{\Lambda^{\alpha}_{.\mu \nu} } $ &
${\phi}_{\mu \nu} \edf C_{\alpha}~\Delta^{\alpha}_{.\mu \nu} $
\\

${\chi}_{\mu \nu} \edf \Lambda^{\alpha}_{. \mu
\nu|{\stackrel{\alpha}{+}} }$ & ${\psi}_{\mu \nu} \edf
\Delta^{\alpha}_{. \mu \nu|{\stackrel{\alpha}{+}}} $
\\

${\varepsilon}_{\mu \nu} \edf C_{\mu | {\stackrel{\nu}{+}}} - C_{\nu
| {\stackrel{\mu}{+}}}$ & ${\theta}_{\mu \nu} \edf C_{\mu |
{\stackrel{\nu}{+}}} + C_{\nu | {\stackrel{\mu}{+}}}  $
\\

${\kappa}_{\mu \nu} \edf \gamma^{\alpha}_{. \mu
\epsilon}\gamma^{\epsilon}_{. \alpha \nu} - \gamma^{\alpha}_{. \nu
\epsilon}\gamma^{\epsilon}_{. \alpha \mu}$   & ${\varpi}_{\mu \nu}
\edf  \gamma^{\alpha}_{. \mu \epsilon}\gamma^{\epsilon}_{. \alpha
\nu} + \gamma^{\alpha}_{. \nu \epsilon}\gamma^{\epsilon}_{. \alpha
\mu}$ \\
 & \\ \hline
 & \\
                  &  ${\omega}_{\mu \nu} \edf \gamma^{\epsilon}_{. \mu \alpha}\gamma^{\alpha}_{. \nu \epsilon}$   \\

                                      &  ${\sigma}_{\mu \nu} \edf \gamma^{\epsilon}_{. \alpha \mu} \gamma^{\alpha}_{. \epsilon \nu}$   \\

                                      &  ${\alpha}_{\mu \nu} \edf C_{\mu}C_{\nu}$   \\

                                      &  $R_{\mu \nu} \edf \frac{1}{2}(\psi_{\mu \nu} - \phi_{\mu \nu} - \theta_{\mu \nu}) + \omega_{\mu \nu}$          \\
 & \\ \hline
\end{tabular}
\end{center}
where $\Lambda^{\al}_{.~\mu \nu |\stackrel{\sigma}{+}} \equiv
\Lambda^{\stackrel{\al}{+}}_{.~\stackrel{\mu}{+}{\stackrel{\nu}{+}}
| \sigma}$. It can be easily shown that there exists an identity
between 2nd order skew-tensors of Table 1, which can be written in the form (cf. Mikhail 1962),
$$\eta_{\mu\nu}+\varepsilon_{\mu\nu}-\chi_{\mu\nu}\equiv
0.\eqno{(2.21)}$$  A useful relation between the torsion and the
contortion is given by Hayashi \& Shirafuji (1979),  $$
\gamma_{\mu\al\nu}=\frac{1}{2}(\Lambda_{\mu\al\nu}-\Lambda_{\nu\mu\al}-\Lambda_{\al\mu\nu})
.\eqno{(2.22)}$$ We see from  Table 1 that the torsion tensor
plays an important role in the structure of AP-space. All tensors in
Table 1 vanish if the torsion vanishes (see (2.22)). The structure
to be used in the present work is characterized by the dual linear
connection (2.5), its corresponding torsion (2.10) and curvature
(2.7) which are, in general, simultaneously non-vanishing objects.
So, it is clear that this structure is more wider than the
Riemannian one and is of the {\it Riemannian-Cartan} type.


\section{ The Field Theory Used}

In this section we are going to review briefly the field equations
of a pure geometric field theory constructed by Wanas \& Ammar (2010),
following a procedure similar to that used in constructing the field
equations of GR, in the context of the geometric structure given in
the previous section. Since this structure is more wider than the
Riemannian one, as mentioned in section 2, it will be shown that the tensor representing the
material distribution is a part of the geometric structure, as expected. The method
used to derive the field equations is that of (Dolan \& McCrea 1963)
variational method \footnote{Since Dolan-McCrea method is not
published, so the reader may refer to  Mikhail \& Wanas (1977), for more
details about its use in the AP-geometry.}is used .  \\

\underline{\bf Field Equations}

 The set of field equations, to be used, has the form Wanas \& Ammar 2010,
$$S^{\mu}_{.~\nu}=0.\eqno{(3.1)}$$
 This set can be written explicitly in the form,
 $$
S^{\mu}_{.~\nu} \edf -2G^{\mu}_{.~\nu}+N \delta^{\mu}_{\nu}-2N^{\mu}_{.~\nu}
+2\gamma^{\gamma\mu}_{.~.~\nu|\stackrel{\gamma}{+}}+2\gamma^{\epsilon\mu}_{.~.~\al}
\gamma^{\al}_{.~\nu\epsilon} $$
$$+\gamma^{\al\mu}_{.~.~\gamma}\gamma^{\gamma}_{.~\al\nu}-\gamma_{\al}^{.~\epsilon\mu}\gamma^{\al}_{.~\nu
\epsilon}-2C_{\al}\gamma^{\al\mu}_{.~.~\nu} = 0,\eqno{(3.2)}$$

where  $N\edf g^{\mu\nu}N_{\mu\nu}$.

To discuss the physical consequences of the geometric set
(3.2), it is convenient to write it in the
covariant form as, \setcounter{equation}{2}
$$
 S_{\nu\sigma}\edf -2G_{\nu \sigma}+N g_{\nu\sigma}-2N_{\nu\sigma}
+2\gamma^{\gamma}_{.~\nu\sigma|\stackrel{\gamma}{+}}+2\gamma^{\epsilon}_{.~\nu\mu}
\gamma^{\mu}_{.~\sigma \epsilon}$$
$$+\gamma^{\al}_{.~\nu\gamma}\gamma^{\gamma}_{.~\al\sigma}+\gamma^{\epsilon}_{.~\mu
\nu}\gamma^{\mu}_{.~\sigma
\epsilon}-2C_{\al}\gamma^{\al}_{.~\nu\sigma}=0,\eqno{(3.3)}
$$
 where $S_{\nu\sigma}\edf g_{\nu\mu}S^{\mu}_{.~\sigma}$.  \\ \\

 \underline{ The Symmetric Part of $S_{\nu\sigma}$}

 The symmetric part of $S_{\nu\sigma}$ is defined as usual by,
$$S_{(\nu\sigma)}\edf\frac{1}{2}(S_{\nu\sigma}+S_{\sigma
\nu}).$$Substituting from (3.3) into the above definition and using
the symmetric tensors of Table (1), we can write
$$S_{(\nu\sigma)}=-2G_{\nu\sigma}-g_{\nu\sigma}\omega
+\psi_{\nu\sigma}+2~\omega_{\nu\sigma}-\phi_{\nu\sigma} = 0,
\eqno{(3.4)}$$ which can be written in the, more convenient, form
$$G_{\nu\sigma}\edf R_{\nu\sigma}-\frac{1}{2}g_{\nu\sigma}R= T^{*}_{\nu\sigma},\eqno{(3.5)}$$
 where,
$$
T^{*}_{\nu\sigma}\edf \frac{1}{2}\psi_{\nu\sigma} -
\frac{1}{2}\varphi_{\nu\sigma}+\omega_{\nu\sigma}-
\frac{1}{2}g_{\nu\sigma}\omega.\eqno{(3.6)}$$  From (3.5) it is
clear that,
 $$ {T^{*}}^{\nu\sigma}_{~~;\sigma}=0.\eqno{(3.7)}$$ This
implies conservation (since the vectorial divergence of the left
hand side of (3.5) vanishes identically). Then the tensor (3.6) can be used to represent the material
 distribution in the theory. \\ \\
\underline{ The Skew-Symmetric Part of $S_{\nu\sigma}$}

The skew part of the field equations (3.5) is given by
$$S_{[\nu\sigma]}\edf\frac{1}{2}(S_{\nu\sigma}-S_{\sigma\nu}) = 0
$$ which can be written, using the skew tensors of Table (1), as
$$S_{[\nu\sigma]}=\chi_{\nu\sigma}-\eta_{\nu\sigma}= 0.\eqno{(3.8)}$$
 Now using the identity (2.21), the skew part of the field
equations can be written in the form,
$$\epsilon_{\nu\sigma}=0, \eqno{(3.9)}$$

\section{Solution With Spherical Symmetry}

The tetrad vector field, which defines the structure of an AP-space
with spherical symmetry, defined by using the coordinate
$(x^{0}\equiv t,x^{1}\equiv r,x^{2}\equiv\theta,x^{3}\equiv\varphi)$
, can be written in the form (Robertson 1932)

$$\h{i}^\mu= \left(%
\begin{array}{cccc}
    A& D r & 0 & 0\\
     0 & B \sin\theta \cos\varphi & \frac{B}{r}\cos\theta \cos\varphi & \frac{-B \sin \varphi}{r \sin \theta} \\
       0 & B \sin\theta \sin\varphi & \frac{B}{r}\cos\theta \sin\varphi  & \frac{B \cos \varphi}{r \sin \theta} \\
       0 & B \cos\theta & \frac{- B}{r}\sin\theta & 0 \\
  \end{array}%
  \right)
\eqno{(4.1)}$$ where $A, B$ and $D$ are function of $r$ only. We are
going to take  $D=0$ which implies the vanishing of all skew tensors
that appear in the structure of the field equations (3.1). This
guarantees the absence of interactions, other than gravity, if any.
In this case (4.1) will reduce to
$$
\h{i}^\mu= \left(%
\begin{array}{cccc}
    A& 0 & 0 & 0\\
     0 & B \sin\theta \cos\varphi & \frac{B}{r}\cos\theta \cos\varphi & \frac{-B \sin \varphi}{r \sin \theta} \\
       0 & B \sin\theta \sin\varphi & \frac{B}{r}\cos\theta \sin\varphi  & \frac{B \cos \varphi}{r \sin \theta} \\
       0 & B \cos\theta & \frac{-B}{r}\sin\theta & 0 \\
  \end{array}%
  \right)
.\eqno{(4.2)}$$ Consequently, using (2.1), we get
$$
\h{i}_\mu= \left(%
\begin{array}{cccc}
    \frac{1}{A}& 0 & 0 & 0\\
         0 &  \frac{1}{B} \sin\theta \cos\varphi & \frac{r}{B}\cos\theta \cos\varphi & \frac{-r}{B} \sin \theta\sin \varphi \\
           0 &  \frac{1}{B} \sin\theta \sin\varphi & \frac{r}{B}\cos\theta \sin\varphi  & \frac{r}{B} \sin\theta \cos\varphi \\
           0 &  \frac{1}{B} \cos\theta & \frac{-r}{B}\sin\theta & 0 \\
  \end{array}%
  \right)
.\eqno{(4.3)}$$ Using definition (2.2) and the tetrad (4.3) we get,
$$
g_{\mu\nu}=\left(%
\begin{array}{cccc}
\frac{1}{A^2} & 0 & 0 & 0  \\
0 & \frac{1}{B^2} & 0 & 0 \\
0 & 0 & \frac{r^2}{B^2} & 0 \\
0 & 0 & 0 & \frac{r^2\sin^2\theta}{B^2}\\
 \end{array}%
 \right)
,\eqno{(4.4)}$$ and then,
$$
g^{\mu\nu}=\left(%
\begin{array}{cccc}
A^2 & 0 & 0 & 0  \\
0 & B^2 & 0 & 0 \\
0 & 0 & \frac{B^2}{r^2} & 0 \\
0 & 0 & 0 & \frac{B^2}{r^2\sin^2\theta}\\
 \end{array}%
 \right)
,\eqno{(4.5)}$$ .\\
From the above tetrad and metric we can evaluate second order tensors of
Table 1, necessary for solving the field equations (3.3), in the
present case.\\ \\
\underline{\it Second Order Symmetric Tensors}

The symmetric tensors necessary for the field equations (3.3),
using the definitions of the tensors given in Table 1, have the following non-vanishing components,
$$
\begin{array}{lll}
&\psi_{00}=2\left[-\frac{B B' A'}{A^3}-B^2\frac{A ''}{A^3}+ B^2\frac{A'^{2}}{A^4}-2\frac{B^2 A'}{A^3r}\right]\\
&\psi_{11}=-4\frac{B'}{B r}\\
&\psi_{22}=2\left[-r^2\frac{B''}{B}-r\frac{B'}{B}\right],
\psi_{33}=\psi_{22}\sin^2\theta
\end{array}
\eqno{(4.6)}$$ and,
$$
\begin{array}{lll}
&\varphi_{00}=B^2\left[-4\frac{A' B'}{A^3 B}-2\frac{A'^{2}}{A^4}\right]\\
&\varphi_{22}=B^2\left[-4\frac{B'^{2}r^2}{B^4} -2 \frac{A'B'r^2}{A
B^3}\right], \varphi_{33}=\varphi_{22}\sin^2\theta~.
\end{array}
\eqno{(4.7)}$$   Also,
$$
\begin{array}{ll}
\omega_{11}=\frac{A'^{2}}{A^2}+2\frac{B'^{2}}{B^2}\\
\end{array}
,\eqno{(4.8)}$$
 and,
$$
\begin{array}{llll}
& R_{00} = \frac{B' A' B}{A^3}- B^2 \frac{A''}{A^3}+ 2B^2
\frac{A'^{2}}{A^4} - 2B^2
\frac{A'}{A^3 r}\\
&R_{11} = -2 \frac{B'}{B r} - 2\frac{B''}{B} - \frac{A''}{A} + 2 \frac{A'^{2}}{A^2} + 2 \frac{B'^{2}}{B^2}-\frac{B'A'}{B A}\\
& R_{22} = -r^2 \frac{B''}{B} - 3r \frac{B'}{B} - r \frac{A'}{A}+ 2r^2 \frac{B'^{2}}{B^2} + r^2 \frac{B' A'}{B A}\\
& R_{33}= R_{22} \sin^2\theta
\end{array}
.\eqno{(4.9)}$$ \\
where $A'\edf\frac{d A}{d r},~~
B'\edf\frac{d B}{d r}$\\ \\
\underline{\it Second Order Skew Tensors}

~~Using the tetrad (4.2), (4.3), the metric(4.4), (4.5) and using the
definitions given in Table(1), we found that all components of the 2nd order skew tensor $\varepsilon_{\alpha \beta}$
vanish identically, as expected. Consequently, the skew part of the field equations, (3.9),
is satisfied identically.\\ \\ \\ \\
\underline{\it Scalars}

~~The scalars which are necessary for the field equations (3.3) can
be obtained from the above components of the second order symmetric
tensors. These are,  $$ \omega = B^2
\left(\frac{A'^{2}}{A^2}+2\frac{B'^{2}}{B^2}\right) \eqno{(4.10)},$$ and
$$ R=2\frac{A' B' B}{A} - 2 B^2 \frac{A''}{A} +
4 B^2 \frac{A'^{2}}{A^2} - 4 \frac{A'}{r A} B^2 + 6 B'^{2}- 4 B'' B
- \frac{8}{r}B' B. \eqno{(4.11)}$$\\
\underline{\it The Field Equations}

 Using the above calculated tensors and the field equations (3.1)
  in its mixed form, we get the following set of
  differential equations,
$$
B^{2}[4\frac{B''}{B}-4\frac{B'^{2}}{B^2}+ 8\frac{B'}{B r} - 2
\frac{A' B'}{A B} + 2\frac{A''}{A} - 3 \frac{A'^{2}}{A^2} + 4
\frac{A'}{A r}] = 0,$$
$$
B^{2}[- 4\frac{B'^{2}}{B^2}+ 8 \frac{B'}{B r} -4 \frac{B'A'}{B A} -
\frac{A'^{2}}{A^2} + 4 \frac{A'}{A r}] = 0,$$
$$
B^{2}[4\frac{B''}{B} - 4 \frac{B'^{2}}{B^2} + 4 \frac{B'}{B r} - 2
\frac{A'B'}{A B} + 2 \frac{A''}{A} - 3 \frac{A'^{2}}{A^2} + 2
\frac{A'}{A r}] = 0. $$ Assuming that $B$ is non-vanishing, to
prevent singularities in (4.3) and (4.4), then we can write the
above set of differential equations in the form, $$
4\frac{B''}{B}-4\frac{B'^{2}}{B^2}+ 8\frac{B'}{B r} - 2 \frac{A'
B'}{A B} + 2\frac{A''}{A} - 3 \frac{A'^{2}}{A^2} + 4 \frac{A'}{A r}
= 0, \eqno{(4.12)}$$
$$
- 4\frac{B'^{2}}{B^2}+ 8 \frac{B'}{B r} -4 \frac{B'A'}{B A} -
\frac{A'^{2}}{A^2} + 4 \frac{A'}{A r} = 0, \eqno{(4.13)}$$
$$
4\frac{B''}{B} - 4 \frac{B'^{2}}{B^2} + 4 \frac{B'}{B r} - 2
\frac{A'B'}{A B} + 2 \frac{A''}{A} - 3 \frac{A'^{2}}{A^2} + 2
\frac{A'}{A r} = 0. \eqno{(4.14)}$$ From (4.12) and (4.14) we get
$$
\frac{A'}{A} = - 2  \frac{B'}{B}, \eqno{(4.15)}$$ which gives, by
integration
$$
A = \frac{C^*}{B^2}, \eqno{(4.16)}$$ where $C^{*}$ is the constant
of integration. The solution (4.16) satisfies equation (4.13)
without any further condition. So, we see that the set of
differential equations (4.12)-(4.14) is not sufficient to
determine the explicit forms of the two function $A$ and $B$. This
will be discussed in section 5.

In what follows we are going to obtain some physical information
from the geometric model given above.

\subsection{A Method for Fixing the Unknown Function}

In this theory, it is clear that the material-energy tensor is a pure
geometric object (3.6), which has the definition (3.6)
$$
T^{*}_{\nu\sigma}\edf \frac{1}{2}\psi_{\nu\sigma} -
\frac{1}{2}\varphi_{\nu\sigma}+\omega_{\nu\sigma}-
\frac{1}{2}g_{\nu\sigma} \omega.$$  Using the second order symmetric
tensors calculated above, we can get the following non-vanishing
components of the "\underline{geometric material-energy tensor}" (3.6),
 $$
T^{*}_{00}= \frac{B^2}{A^2}\left[\frac{A'B'}{A B} - \frac{A''}{A} +
\frac{3}{2}\frac{A'^{2}}{A^2}-2\frac{A'}{A r} -
\frac{B'^{2}}{B^2}\right], \eqno{(4.17)}$$
 $$
T^{*}_{11}=\frac{B'^{2}}{B^2} - 2 \frac{B'}{B r} + \frac{A'^{2}}{2
A^2}, \eqno{(4.18)}$$
$$
T^{*}_{22} = r^2  \left[-\frac{B''}{B} + \frac{B'^{2}}{B^2} -
\frac{B'}{B r} + \frac{A'B'}{A B} - \frac{A'^{2}}{2 A^2}\right]
,\eqno{(4.19)}$$ and
$$
T^{*}_{33} = T^{*}_{22} \sin^2\theta .$$ Using the definition
${T^{*}}^{\mu}_{.~\nu} \edf g^{\al\mu} T^{*}_{\al\nu}$ ,(4.15) and
the metric (4.5) we can write the above components in the form, $$
{T^{*}}^{0}_{.~0} = B^2 \left[-3\frac{B'^{2}}{B^2} + 2 \frac{B''}{B}
+ 4 \frac{B'}{B r}\right], \eqno{(4.20)}$$
$$
 {T^{*}}^{1}_{.~1} = B^2 \left[ 3 \frac{B'^{2}}{B^2} - 2
\frac{B'}{B r}\right], \eqno{(4.21)}$$
$$
{T^{*}}^{2}_{.2} = B^2 \left[ - \frac{B''}{B} - \frac{B'}{B r} - 3
\frac{B'^{2}}{B^2}\right], \eqno{(4.22)}$$ and
$${T^{*}}^{3}_{.~3}={T^{*}}^{2}_{.~2}~~.$$
 If we assume that we have
a {\it perfect fluid}, then the components of the  geometric energy-momentum tensor will
satisfy the following relation
$${T^{*}}^{1}_{.~1}={T^{*}}^{2}_{.~2}={T^{*}}^{3}_{.~3}~~.\eqno{(4.23)}$$
Then, using this relation, i.e. equating (4.21) and (4.22), we get
$$ \frac{B''}{B}+6\frac{B'^2}{B^2}-\frac{B'}{B r}=0~,$$ which is a second order differential
 equation in $(B)$ only. Integrating
this differential equation twice we get
$$B = (7C\frac{r^2}{2} + 7 C_1)^{\frac{1}{7}}.\eqno{(4.24)}$$\\
So, from the relation (4.16) we can write
$$
A = \frac{C^*}{B^2}=\frac{C^*}{(7C\frac{r^2}{2} + 7
C_1)^{\frac{2}{7}}}~ .\eqno{(4.25)}$$ This shows that the field
equations (3.3) can fix the unknown functions of the model, only if
we attribute some properties to the material distribution. \\

\subsection{ The Metric of the Associated Riemannian Space}

In order to get some physical information about the solution
obtained, (4.24) and (4.25), we are going to write the metric of
Riemannian space, associated with the AP-space (4.3), which is
given, in general, by
$$d s^2=g_{\mu\nu} d x^{\mu} d x^{\nu}.\eqno{(4.26)}$$ Substituting
the solution obtained, (4.24) and (4.25), into (4.4) we can write,
$$
\begin{array}{ll}
ds^2 =& \frac{1}{(C^*)^2}(7C\frac{r^2}{2} + 7 C_1)^{\frac{4}{7}}dt^2\\
&+\frac{1}{(7C\frac{r^2}{2} + 7
C_1)^{\frac{2}{7}}}(dr^2+r^2d\theta^2+r^2\sin^2\theta d\varphi^2)
\end{array}
,\eqno{(4.27)}$$ in order to facilitate comparison with the results
of GR, we are going to write (4.27) assuming that $(C^*)^2=-C_2$ (
where $C_2$ is another arbitrary constant) and $d\tau= i ds$, then
we get,
$$
\begin{array}{ll}
d\tau^2 =& \frac{1}{C_2}(7C\frac{r^2}{2} + 7 C_1)^{\frac{4}{7}}dt^2\\
&-\frac{1}{(7C\frac{r^2}{2} + 7
C_1)^{\frac{2}{7}}}(dr^2+r^2d\theta^2+r^2\sin^2\theta d\varphi^2)
\end{array}
,\eqno{(4.28)}$$ which is the metric of the pseudo-Riemannian space
associated with the AP-space, with the same signature used usually
in such applications of GR. \\ \\
\underline{\it Boundary Conditions}

~~The non-vanishing of some of the components of the geometric
material-energy tensor (3.6), shows clearly that the solution
obtained is an interior one, having spherical symmetry. The field equation (3.3), for the present geometric structure (4.2), will reduce to Einstein field equations ,
$$
R_{\mu\nu}=0,
$$
if the geometric material-energy tensor (3.6) vanishes. So, the solution of (3.3) in this case would be given by the Schwarzschild exterior solution. To fix the
constants of integration $C, C_1$ and $C_2$, we are going to make a
matching between the metric (4.28) and the Schwarzschild exterior
metric (cf. Adler et al. 1975) at the boundary $ r=a$. Now, let us define the geometric
density $({\rho^{*}}_{0})$ and the geometric pressure
$({p^{*}}_{0})$ (these quantities are in relativistic
units), we get
$$
{\rho^{*}}_{0}(r)\edf -{T^{*}}^{0}_{.~0} \quad , \quad {p^{*}}_{0}(r) \edf {T^{*}}^{1}_{.~1} =
{T^{*}}^{2}_{.~2} = {T^{*}}^{3}_{.~3} .
\eqno{(4.29)}$$  Then we can determine ~ $p^{*}_{0}$ ~ from equation
(4.21),(4.29) using the solution (4.24) and (4.25). This can be
written in the form,
$$ p^{*}_{0}(r)=\frac{(4 C^2 r^2 + 14 CC_1)}{(7C\frac{r^2}{2} + 7
C_1)^{\frac{12}{7}}}. \eqno{(4.30)}$$ Then, assuming that
$p^{*}_{0}(r)=0$ at the boundary  $r=a$ we get

\[
C=-~\frac{7C_1}{2a^2}~,
\]
and from matching the metric (4.28) with the Schwarzschild exterior
solution, which can be written in the form (cf. Adler et al. 1975),
 $$ d\tau^2
=\frac{(1-\frac{m}{2 r})^2}{(1+\frac{m}{2 r})^2}dt^2 - (1+\frac{m}{2
r})^{4}(dr^2+r^2d\theta^2+r^2\sin^2\theta d\varphi^2) ,$$ at $r=a$
we get,
\[
7 C_1 = (-~\frac{4}{3})\frac{1}{(1+\frac{m}{2a})^{14}}~,
\]
and
\[
C_2=\frac{1}{(1-\frac{m}{2a})^{2} (1+\frac{m}{2a})^{6}}~.
\]
Consequently we can write the solution (4.24) and (4.25) in the
form,
$$B=\frac{(\frac{-4}{3})^{\frac{1}{7}}(1-\frac{7 r^2}{4
a^2})^{\frac{1}{7}}}{(1+\frac{m}{2 a})^{2}}~,\eqno{(4.31)}$$ and
$$A=\frac{i (1+\frac{m}{2 a})}{(1-\frac{m}{2 a})(\frac{-4}{3})^{\frac{2}{7}}
(1-\frac{7 r^2}{4 a^2})^{\frac{2}{7}}}~.\eqno{(4.32)}$$ Then the
metric (4.28) can now be written in the form,
$$
d\tau^2=b_{1} (1-\frac{7}{4a^2}r^2)^{\frac{4}{7}}dt^2~-~\frac
{b_{2}}{(1-\frac{7 r^2}{4
a^2})^{\frac{2}{7}}}(dr^2+r^2d\theta^2+r^2\sin^2\theta d\varphi^2)~,
\eqno{(4.33)}$$ where $$b_{1}\edf(\frac{-4}{3})^{\frac{4}{7}}
\frac{(1-\frac{m}{2a})^{2}} {(1+\frac{m}{2a})^{2}},\eqno{(i)}$$
and
$$b_{2}\edf\frac{(1+\frac{m}{2a})^{4}} {{(\frac{-4}{3})}^{\frac{2}{7}}}.\eqno{(ii)}$$
Substitute from the solution (4.31) into (4.20) and (4.21) using $(i)$ and $(ii)$
, we can evaluate
 $p^{*}_{0}(r)$ and $\rho^{*}_{0}(r)$ in the forms,
$$
p^{*}_{0}(r) = \frac{1}{a^2 b_{2}} \frac{
(1-\frac{r^2}{a^2})}{(1-\frac{7}{4a^2}r^2)^{\frac{12}{7}}}
~,\eqno{(4.34)}$$ and
$$
\rho^{*}_{0}(r) = 3~ \frac{1}{a^2 b_{2}}
\frac{(1-\frac{r^2}{2a^2})}{(1-\frac{7}{4 a^2}r^2)^{\frac{12}{7}}}
~.\eqno{(4.35)}$$
These two equations give the values of the geometric pressure and density as
functions of the radial distance $r$.

\section{Discussion}

~~~In the present work, we have applied a pure geometric field
theory, in which the tensor representing the material-energy
distribution, given by (3.6), is a part of the geometric structure
used. The field equations of the theory are applied to a geometric
structure having spherical symmetry. It appears that the field
equations are not sufficient to fix the unknown functions in this
case. The non-vanishing components of the geometric
material-energy tensor, as clear from equations (4.17), (4.18) and
(4.19), shows that the model represents the field within a material
distribution. It is shown that, the unknown functions of the model can
be fixed by assuming, in addition to the field equations, that a
perfect fluid is filling the spherical configuration. Certain components of the material-energy tensors (4.29) are chosen to represent geometric
density and pressure. Some boundary conditions
are used to fix the constants of integration. This is done by
assuming, at the boundary $(r=a)$, that:

(i) The geometric pressure $(p^{*}_{0})$ vanishes.

 (ii) The metric (4.28) obtained from the solution and the Schwarzschild
 exterior metric should be the same at this boundary.\\ \\
Now, we  direct the attention to the following points:\\ \\
(1) {\bf The field theory} used can be classified as a pure geometric
one. In other words, all physical quantities are represented by
geometric objects that are constructed from the building blocks of
the geometric structure used, an AP-structure of the {\it
Riemann-Cartan} type. This is done to avoid the use of a
phenomenological material-energy tensor and an equation of state.\\
The theory proposed reduces to GR outside the material distribution, i.e when
$ T^{*}_{\mu\nu}$(see equation $(3.5), (3.6))$. So, it will give all the results of GR in the Solar
system. In other words, the theory is in good agreement with the results of laboratory
and Solar system tests. We stress here on the fact that the type of AP-geometry used in the present work is
different from that used in the literature. In the case of $T^{*}_{\mu\nu}\neq
0$
, there is no necessity to get GR results, since, on one hand most of the problems of
applications of GR emerge from the use of the field equations $(1.1)$
with a phenomenological $T_{\mu\nu}$ as mentioned in the introduction. On the other
hand  $T^{*}_{\mu\nu} $ in the present work is a part of the geometric structure used.
 \\ \\
(2) {\bf The geometric structure} with spherical symmetry, given in
section 3, used in the present work, is that usually used in such
applications (cf. Wanas 1985, 2007). The non-vanishing of 2nd-order skew
tensors (see Table 1) may give rise to interactions other than
gravity. So, to restrict the model to gravitational interactions
only, we have assumed the vanishing of the function $D$ which leads to the
vanishing of 2nd order skew tensors, relevant to the theory used. \\
\\
(3) It is shown that {\bf the field equations} (3.1) are not
sufficient to determine the unknown functions $A(r)$, $B(r)$ of the
model. Instead, it gives a relation (4.16) between $A$ and $B$. Any
pair of functions satisfying this relation represents a solution.
This means that relation (4.16) implies a family of solutions. On the other
hand, the spherically symmetric geometric structure, used in the
present study, can be considered as admitting representation of
several physical situations. Once one of such situations (perfect
fluid) is assumed, the functions can be fixed as shown in (4.31),
(4.32). This has to be expected from the beginning, since the
components of the material-energy tensor (4.20), (4.21) and (4.22),
are pure geometric objects, without imposing any restrictions, expect for spherical
symmetry. In other words, the relation (4.16) represents several physical situations, including perfect fluid. Once a certain situation is fixed, the field equations can be solved complectly. \\ \\
(4) {\bf The boundary condition} used in the previous section, $r=a$,
can be considered as a boundary between two regions: The
first is characterized by $p^{*}_{0}\neq 0$ while the second
is characterized by $p^{*}_{0}= 0$. We can consider the value $(a)$ as
representing the radius of the spherical configuration ( a static or a slowly rotating star). This boundary condition can be applied to Sun-like-stars. The pressure near the center of the Sun likes stars is much more higher than that in the Corona. So, this pressure can be neglected at the boundary relative to its value at the center, in such type of stars.

The radial distance $(r)$, measured from the center of the star, can be written in terms of the
dimensionless quantity,
$$q\edf\frac{r}{a}~.$$ In this case the solution (4.31) and (4.32), the geometric
pressure (4.34) and the geometric density (4.35) can be written in
terms of $q$ and the parameters $b_{1}$ and $b_{2}$, of the previous section, as
$$B=\frac{(1-\frac{7 }{4} q^2)^{\frac{1}{7}}}{\sqrt{b_{2}}}~,\eqno{(5.1)}$$
$$A=\frac{i}{\sqrt{b_{1}}
(1-\frac{7}{4} q^2)^{\frac{2}{7}}}~,\eqno{(5.2)}$$

$$
p^{*}_{0}(r) = \frac{1}{a^2 b_{2}} \frac{
(1-q^2)}{(1-\frac{7}{4}q^2)^{\frac{12}{7}}} ~,\eqno{(5.3)}$$ and
$$
\rho^{*}_{0}(r) = 3 \frac{1}{a^2 b_{2}}
\frac{(1-\frac{q^2}{2})}{(1-\frac{7}{4}q^2)^{\frac{12}{7}}}
~.\eqno{(5.4)}$$
\\ \\
(5) {\bf Corona}: If we evaluate the geometric density (4.35) at
$r=a$, or $q=1$, we get
$$\rho^{*}_{0}=\frac{0.7771717}{a^2 b_{2}}\eqno{(5.5)}.$$
If we consider the model obtained as representing a certain type of main sequence stars, then imposing the condition for vanishing density $(\rho^{*}_{0}=0)$ on
(5.4), we get from (4.35) $$q=\sqrt{2}.\eqno{(5.6)}$$ So, we have a
region (spherical shell), outside the star, within which
$p^{*}_{0}=0$ while $\rho^{*}_{0}\neq 0$. The width of this region
is in the range $\sqrt{2}>q\geq 1$. This region can be considered as a
corona of the star. However, the corona of the Sun is not an exact
spherical shell. This is due to the magnetic activity of the Sun.
Recall that the model in the present treatment is
a pure gravity model, i.e. without considering any electromagnetic influence. This may
give an interpretation for the difference between the shape of the corona, given in
the present work, and that of an actual star. \\ \\
(6) {\bf Equation of state}: Using the definitions of the geometric
pressure (5.3) and density (5.4), we can write the following
relation between $(p^{*}_{0})$ and $(\rho^{*}_{0})$, as
$$p^{*}_{0}=\frac{(1-q^2)}{3(1-\frac{q^2}{2})}~~
\rho^{*}_{0}~.\eqno{(5.7)}$$ This represents a general {\it equation of state}, for perfect fluids,  that is
obtained from ,and not imposed on, the model. Equation (5.7) is one of the
advantages of using a pure geometric theory of gravity. In fact
equation (5.7) represents a continues spectra of characteristics of
the geometry induced material distribution within the star. This spectra can not be
obtained if one uses the conventional approach, since the equation
of state is imposed on the model, in that case.

Fig.1 gives the relation between the two dimensionless quantities
 $(q)$  and $(\frac{{p^{*}}_{0}}{{\rho^{*}}_{0}})$.

     \begin{figure}
   \centering
   \includegraphics[width=4cm]{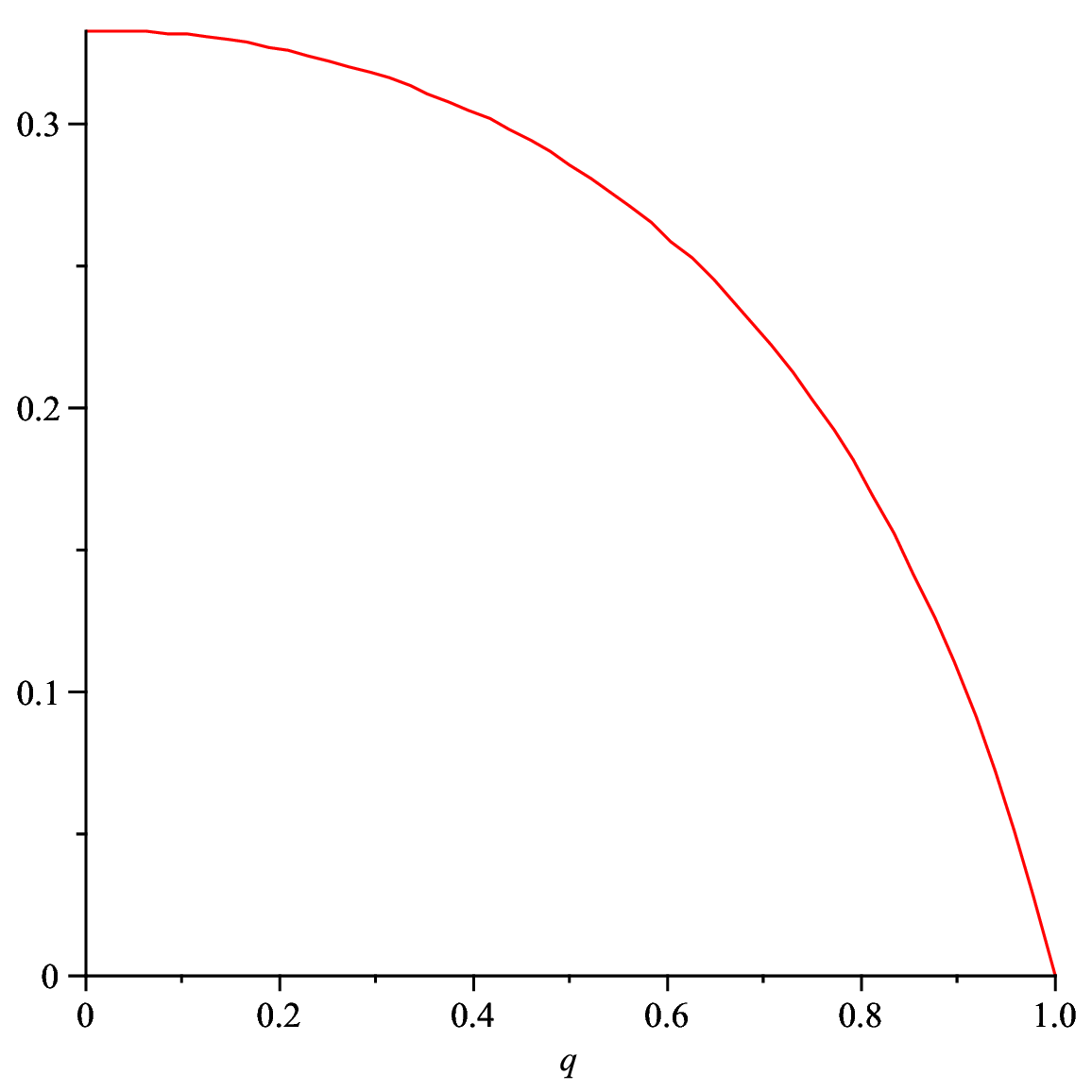}
      \caption{The relation between $q$ and $\frac{p^{*}_{0}}{\rho^{*}_{0}}$
              }
         \label{FigVibStab}
   \end{figure}

The relation between the geometric material-energy tensor and the
phenomenological one (written in cgs unit) is obtained by comparing
the R.H.S. of (3.5) and the R.H.S. of (1.1) so, we can write
$$
{T^{*}}^\mu_{~\nu}=-\kappa ~ T^\mu_{~\nu}. \eqno{(5.8)}$$  Then we
get, $$\frac{p^{*}_{0}}{\rho^{*}_{0}}=\frac{p_{0}}{\rho_{0}} ~,$$
where $p_{0}$ and $\rho_{0}$ are the phenomenological pressure and density
of the material distribution, respectively. The above relations, (5.7) and (5.8), and
Figure 1 show how
 $(\frac{p_{0}}{\rho_{0}})$ varies from the
center to the surface of the star.\\ \\
(7) {\bf Radiation Zone}: It is clear from the equation of state
(5.7) (or from Fig.1) that as $q\Rightarrow 0$,
$\frac{p_{0}}{\rho_{0}}\Rightarrow \frac{1}{3}$. This value
characterizes the situation of a radiation dominant central zone.
This situation washes out gradually as we move towards the surface
 of the star $(q=1)$ , at which $p_{0}=0$, which characterizes
 energy transfer in a class of main sequence stars (cf. Abhyankar 1992), according to Schwarzschild classification. \\ \\
(8) {\bf Singularities and Convection Zone}: It is clear that the
solution given (5.2) is singular at $q=\sqrt{\frac{4}{7}}\simeq
0.755928$. It is also clear from that (5.3)) and (5.4) that the geometric
pressure and density are also singular at the same value of $q$ ,
while the equation of state (5.7) is still regular.

The appearance of singularity, may indicate that the model is
applied outside its domain of applicability. In other words, the
model may include one or more assumptions that is not applicable at
the singularity. Let us recall the case of the Sun. There is a
convection zone at a depth about $28\%$ of the radius of the Sun,
i.e. $72\%$ of the radius as measured from the center of the Sun
(cf. Tayler 1994). The perfect fluid assumption within the convection zone is no
longer valid. The presence of singularity in the present work may
indicate the existence of a zone in which perfect
fluid assumption is violated, a {\it convective zone}. Fig.2 shows the
variation of the density and pressure as function of $q$.\\  \\
(9) {\bf The model obtained}, in the present work, is far from
representing a compact object. This is because the equation
    \begin{figure}
   \centering
   \includegraphics[width=4cm]{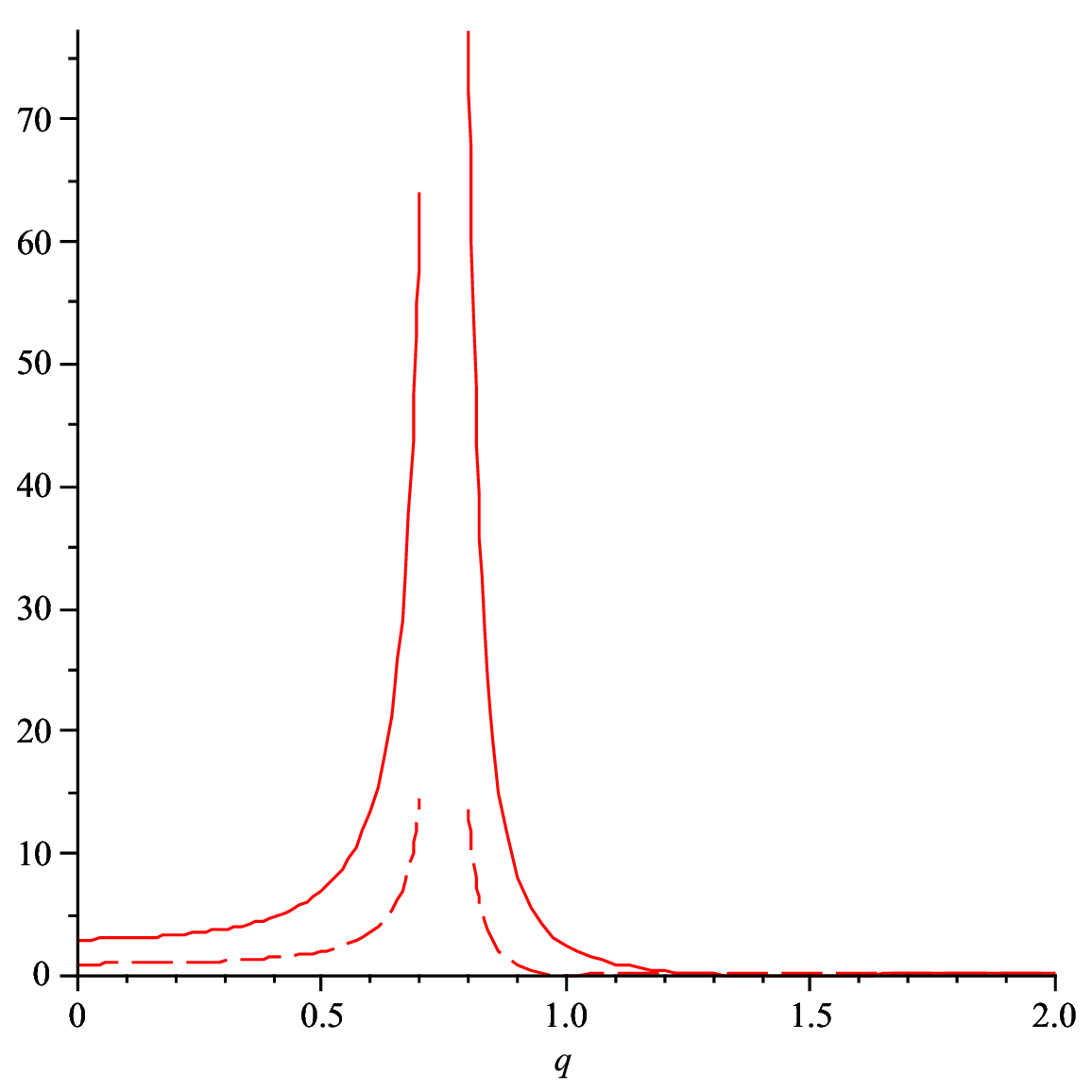}
      \caption{The relation between $q$ and $p^{*}_{0}$ is represented by the dashed
line, while the relation between $q$ and ${\rho^{*}}_{0}$ is
represented by the solid line.
              }
         \label{FigVibStab}
   \end{figure}
 of state for such objects is the adiabatic one  (cf. Shapiro \& Teukolsky 1983), given by
$$p_{0}=K \rho^{1+\frac{1}{n}}_{0}~,\eqno{(5.9)}$$ where
$(K)$ is constant and $(n)$ is the polytropic index. This index
varies from $(\frac{3}{2})$ for non-relativistic particles to $(3)$
for relativistic particles. It is clear that the equation of state
obtained, (5.7), has a polytropic index tending to infinity, which
characterizes an isothermal configuration (cf. Abhyankar 1992).

\section{Concluding Remarks}

Many of gravity theories suggested have satisfactory PN-parameters, but
they are not capable of covering the results of recent observation, e.g.
SN-type Ia, the velocity curves of spiral galaxies, ...  . The reason may be the use
of phenomenological material-energy tensor. The present treatment throws some light on this problem.

The situation here is similar to the Schwarzschild exterior solution in which pure geometric
considerations led to many successful physical interpretation and predictions.
This shows that Einstein geometerization scheme can be extended to cover domains
with material distributions.

We may conclude that the model obtained in the present work
represents a lower main sequence star according to the Schwarzschild classification  (cf. Abhyankar 1992) with a corona, radiation zone
and convection zone.

As far as we know, this is the first time to obtain the above mentioned physical results from pure geometric considerations. This  deserves consideration, since the model obtained in the present treatment deals
with a region (stellar interior)for which matter has no direct observation.

We would like to point out that the introduction of electromagnetic effects, on the model,
may throw more light on the physical situation within the star. This can be done by relaxing the condition,
$D(r)=0$, imposed on the model.

Finally, it worth of mention that the model obtained, in the present work, is far from being a complete one for a realistic star. It is just an attempt to construct a permeative model from pure geometric consideration. It can be considered as a first step in a series, which needs more efforts for other steps in this direction.  The importance of this approach is that stellar interiors are not accessible by direct observation, so far.

\section*{Acknowledgements}

The authors would like to thank members of the " Egyptian Relativity Group", for many discussions. In particular, they
would like to thank professor G.G.L.Nashed and Miss Mona M. Kamal, for checking some of the calculations.

\section*{References}
Abhyankar K.D., 1992,   Astrophysics: Stars and Galaxies, Tate McGraw-Hill.\\
Adler R., Bazin M., Schiffer, M., 1975, Introduction
to General Relativity, 2nd.ed.,

McGraw-Hill, New York.\\
Dolan P., McCrea W. H., 1963, Personal Communications.\\
Einstein A., 1955, The Meaning Of Relativity, 5th ed.
priceton.\\
Hayashi K., Shirafuji T., 1979, Phys. Rev. D, 19,
3524.\\
Mikhail F.I., 1962, Ain Shams Sci. Bul., 6, 87.\\
Mikhail F.I., Wanas M.I., 1977, Proc. Roy. Soc. Lond.
A., 356, 471.\\
Robertson H.p., 1932, Ann. Math., Princeton (2),
 33, 496.\\
Shapiro S.L., Teukolsky S.A. ,1983, Black Holes, White
Dwarfs and Neutron Stars,

John Wiley and Sons.\\
Tayler R.J., 1994,  The stars: their structure and
evolution, 2nd ed.

Cambridge University.\\
Wanas M.I., 1985, Int.J.Theor. Phys., 24, 639.\\
Wanas M.I., 2001, Stud. Cercet. Stiin. Ser. Mat.Univ.Bacau, 10, 297;

preprint (gr-qc/0209050).\\
Wanas M.I., 2007, Int. J.Geom.Methods.Mod. Phys., 4, 373; preprint (gr-qc/ 0703036).\\
Wanas M.I., Ammar S.A., 2010, Mod. Phys. Lett. A., 25, 1705.
\end{document}